# General model of pressure-induced transition temperature increase with focus on the Hg-Ba-Ca-Cu-O System

Mario Rabinowitz[a] and T. McMullen[b]

[a]*Electric Power Research Institute, Palo Alto, CA 94303 USA lrainbow@stanford.edu*

[b]*Department of Physics, Virginia Commonwealth University, Richmond, Virginia 23284*

**Abstract**

A general calculation of high pressure induced transition temperature ($T_c$) increases for oxide superconductors gives a good description for the Hg-Ba-Ca-Cu-O system. Predictions are made of the maximum $T_{cP}^{theo} \approx$ 163, 154, and 111K compared with $T_{cP}^{exp}$ = 164, 154, and 118K respectively for the superconducting phases of Hg-1223, Hg-1212, and Hg-1201 as a function of their compressibilities agreeing with experiment. The theoretical result is in the form of a universal scaling equation for the fractional change in $T_c$ as a function of pressure, which is derived from a recent phenomenological theory of short coherence length superconductivity.

## 1. Introduction

We have followed with great interest the recent exciting work in which the transition temperature ($T_c$) of $HgBa_2Ca_2Cu_3O_{8+\delta}$ has been increased in a matter of months from a record high ≈ 133.5 K [1] to 135 K [2] at atmospheric pressure, and to another record high 164 K by the application of quasihydrostatic high pressure. Two independent research groups [2,3] concurrently reported superconductivity at $T_c$ > 153 K by application of pressure ≈ 23 GPa in this system. More recently the TCSUH group [4] have taken the optimally doped $HgBa_2Ca_{q-1}Cu_qO_{2q+2+\delta}$ (where q = 1, 2, and 3) system up to ≈ 164K for Hg-1223, ≈ 154K for Hg-1212, and ≈ 118K for Hg-1201.

The increase of $T_c$ with increasing pressure, passing through a broad peak, has been thought to be due to a pressure-induced change in carrier concentration [5, 6]. A generally successful working hypothesis is that application of high pressures < 20 GPa produces an increase in $T_c$ for underdoped cuprates, whereas this leads to a decrease in $T_c$ for overdoped cuprates. A negligible effect is expected for optimal doping on maximum $T_c$ , making the increases for all three phases of the optimally doped Hg-Ba-Ca-Cu-O system anomalous and hence all the more remarkable. A recent high pressure observation of $YBa_2Cu_3O_7$ also makes it likely that additional unknown effects are operational [7]. It is clear that these effects can not be explained in terms of former considerations. Neumeir and Zimmerman [7] make an empirical fit to their data in developing a phenomenological theory of the pressure effect on $T_c$.

We were motivated to write this paper for several reasons. One is that a predictive theory of high pressure-induced $T_c$ increase is needed. A second reason is it presents another test of a highly successful non-interactive model of short coherence length superconductivity [8,9,10]. Another is to shed light on the anomalous aspect of the new high pressure results in $YBa_2Cu_3O_7$ and the Hg-Ba-Ca-Cu-O system. Finally, the results with the Hg-Ba-Ca-Cu-O system are so impressive that they present an important challenge for explanation.

**2. Analysis**

As derived from a phenomenological theory of superconductivity [8,9,10], the transition temperature for short coherence length materials is

$$T_c = A\left[\frac{\hbar^2 n^{2/3}}{2mk}\right], \qquad (1)$$

where $A = A_3 = 0.218$ for isotropic three--dimensional (3-D) superconductors and $A_2 = (3/2)A_3 = 0.328$ for anisotropic 2-D superconductors, $\hbar$ is (Planck's constant)/$2\pi$, m is the carrier effective mass, and k is the Boltzmann constant. A nice aspect of the present analysis is that it results in a test of the general scaling wherein the specific value of A does not enter in, as A cancels out.

The number density of carriers is n = N/V, where N is the number of carriers and V is the volume of the sample. Taking the partial derivative of $T_c$ in eq. (1) with respect to pressure P yields

$$\frac{\partial T_c}{\partial P} = \frac{2}{3}T_c\left[-\frac{1}{V}\frac{\partial V}{\partial P} + \frac{1}{N}\frac{\partial N}{\partial P}\right] - \frac{T_c}{m}\frac{\partial m}{\partial P}\ . \qquad (2)$$

The inverse of the bulk modulus is the compressibility $\kappa = -\frac{1}{V}\frac{\partial V}{\partial P}$, so eq. (2) can be written as a universal scaling equation for the fractional change in $T_c$:

$$\frac{1}{T_c}\frac{\partial T_c}{\partial P} = \frac{2}{3}\left[\kappa + \frac{1}{N}\frac{\partial N}{\partial P}\right] - \frac{1}{m}\frac{\partial m}{\partial P}\ . \qquad (3)$$

Thanks to the previously developed extraordinarily successful Cornelius and Schilling model [11], we have estimates of the compressibility $\kappa$ for $HgBa_2Ca_{q-1}Cu_qO_{2q+2+\delta}$ of (1/88) $(GPa)^{-1}$, (1/94) $(GPa)^{-1}$, and (1/101) $(GPa)^{-1}$ with q = 1, 2, and 3 respectively. They assert that their model estimates the bulk modulus of the oxide superconductors with a < 8% error relative to experimentally determined values. Our estimates are based upon their $\kappa$ values.

Since the variations $\partial N/\partial P$ and $\partial m/\partial P$are not known, let us neglect them to a first approximation at this time. We expect $\partial N/\partial P$ to be small, and will examine $\partial m/\partial P$ in the section 3. Thus we are here determining the extent to which the compressibility alone contributes to the high pressure increase of $T_c$ from calculations based upon eq. (3). The results for $\frac{1}{T_c}\frac{\partial T_c}{\partial P}$ are shown in Table 1 with the theoretical values comparing well with the mean experimental values of Gao et al [4] where for Hg-1201 $T_c$ goes from 94K to 118K with a pressure increase of ≈ 24 GPa; where for Hg-1212 $T_c$ goes from 128K to 154K with a pressure increase of ≈ 29 GPa; and where for Hg-1223 $T_c$ goes from 135K to 164 K with a pressure increase of ≈ 31 GPa.

Using the known $T_c$ values at atmospheric pressure, we find $\frac{\partial T_c}{\partial P}$ which compares well with the mean experimental values of Gao et al [4] as shown in Table I. This enables us to calculate the high pressure maximum transition temperature $T_{cP}^{theo}$ in terms of the atmospheric pressure transition temperature $T_c^{exp}$ and κ. Integrating eq. (3):

$$T_{cP}^{theo} \approx T_c^{exp} [1 + (2/3)\kappa(\Delta P)],$$

(4)

where ΔP is the pressure increase. Table I shows excellent agreement between $T_{cP}^{theo}$ and $T_{cP}^{exp}$. The 6% lower $T_{cP}^{theo}$ for Hg-1201 may be indicative that the peak value of $T_{cP}^{exp}$ occurs at 26 GPa rather than 24 GPa which would give $T_{cP}^{theo}$ = 118 K. This is a possibility as there are far fewer data points for this phase than the other two phases. There is also a larger error for this phase for the experimental points relative to the smooth curve fit to the data.

**3. Effective Mass Contribution**

We will here look into a possible contribution from a potential variation of the effective mass with pressure. There are many kinds of effective mass. The effective mass of a carrier we shall consider is given by

$$m_i = \hbar^2 / \left(\partial^2 E / \partial k_i^2\right), \tag{5}$$

where the wave vector $k_i = p_i / \hbar$, and $p_i$ is the momentum in the i direction. The energy dispersion $E(\mathbf{k})$ can be complicated so that m may either decrease, stay constant, or increase depending on the crystal lattice and how the band structure is affected by pressure.

To qualitatively examine the effects of pressure on the effective mass, let us consider a simple cubic lattice. The energy of an s electron in the crystal in the tight binding model may be found in Dekker [11]:

$$E(\mathbf{k}) = E_o - \alpha - 2\gamma(\cos k_x a + \cos k_y a + \cos k_z a), \tag{6}$$

where a is the edge length of the unit cell, $E_o$ is the energy of the electron in the free atom, $\alpha$ and $\gamma$ are related to the mean potential energy of an electron at a given point resulting from the presence of all atoms except the atom located at that point. This is given by Dekker [11] in integral form in terms of the wave functions corresponding to nearest neighbors. For our purpose, it is sufficient to know that $\gamma$ is positive. Since the cosine terms vary between $\pm 1$, the total band width is $12\gamma$. The band width may be expected to increase with increasing pressure since as atoms are forced closer together, the overlap of the wave functions of neighboring atoms increases. That is why outer electron levels give rise to wider bands than the inner shell electrons.

Expanding the cosine terms to second order for small k,

$$E(\mathbf{k}) \approx E_o - \alpha - 6\gamma + \gamma a^2 k^2 \tag{7}$$

Applying eq. (5) to eq. (7) determines the effective mass,

$$m = \hbar^2 / (2\gamma a^2). \tag{8}$$

Thus as $\gamma$ increases with pressure due to increasing band width, eq.(8) indicates that to first approximation m decreases inversely proportional to $\gamma$. This is equivalent to the hopping matrix prescription that the effective mass decreases as the hopping matrix increases, and the perspective that as electrons are more weakly bound they have a smaller effective mass allowing them to move more easily between scattering events. Of course, these qualitative notions are superseded by the second derivative of the energy dispersion as given by eq. (5) so that in more complicated lattices, m may either decrease, stay constant, or increase with increasing pressure.

For our simple example, substituting eq. (8) into eq. (3):

$$\frac{1}{T_c}\frac{\partial T_c}{\partial P} = \frac{2}{3}\left[\kappa + \frac{1}{N}\frac{\partial N}{\partial P}\right] + \frac{1}{\gamma}\frac{\partial \gamma}{\partial P} \; . \tag{9}$$

Thus we see qualitatively that our calculation of the effects of high pressure on the transition temperature can be increased or decreased by the term related to the change in effective mass with pressure as the band width changes. In more complicated systems this term can in principle be more dominant than it appears to be in the Hg-Ba-Ca-Cu-O system.

## 4. Conclusion

We have shown that by using the Rabinowitz model of short coherence length superconductivity[8, 9, 10], it is possible to predict the high pressure maximum transition temperature $T_{cP}$, $\frac{\partial T_c}{\partial P}$, and $\frac{1}{T_c}\frac{\partial T_c}{\partial P}$ quite accurately using only the compressibility for the Hg-Ba-Ca-Cu-O system. It remains to be determined from a more accurate calculation for other systems, whether inclusion of the increase related to effective mass is necessary to account for the the experimental values. It appears to be unnecessary for this system up to the maximum value of $T_{cP}$. Our results are general and should apply to all short coherence length superconductors.

TABLE I. Comparing Experimental and Theoretical Values

| Compound | $T_c^{exp}$ (K) | $T_{cP}^{exp}$ (K) | $T_{cP}^{theo}$ (K) | $\left\langle \frac{1}{T_c}\frac{\partial T_c}{\partial P}\right\rangle^{exp}$ $(GPa)^{-1}$ | $\left[\frac{1}{T_c}\frac{\partial T_c}{\partial P}\right]^{theo}$ $(GPa)^{-1}$ | $\left\langle \frac{\partial T_c}{\partial P}\right\rangle^{exp} \left(\frac{K^o}{GPa}\right)$ | $\left[\frac{\partial T_c}{\partial P}\right]^{theo} \left(\frac{K^o}{GPa}\right)$ |
|---|---|---|---|---|---|---|---|
| Hg 1201 | 94 | 118 | 111 | $1.06 \times 10^{-2}$ | $7.58 \times 10^{-3}$ | 1.00 | 0.713 |
| Hg 1212 | 128 | 154 | 154 | $7.01 \times 10^{-3}$ | $7.09 \times 10^{-3}$ | 0.897 | 0.908 |
| Hg 1223 | 135 | 164 | 163 | $6.93 \times 10^{-3}$ | $6.60 \times 10^{-3}$ | 0.935 | 0.891 |

**References I didn't use. Save for another paper.**